\documentclass[a4paper]{jpconf}
\usepackage{graphicx}
\usepackage[utf8]{inputenc}
\usepackage[T1]{fontenc}
\usepackage{enumitem}
\usepackage{mathptmx}
\usepackage[svgnames]{xcolor} 
\usepackage{hyperref}         
\usepackage{amsmath}          
\usepackage{amssymb}          
\usepackage{amsfonts}         
\usepackage{amstext}          
\usepackage{textcomp}         
\usepackage{slashed}
\usepackage{cite}

\DeclareMathAlphabet{\mathcal}{OMS}{cmsy}{m}{n}

\renewcommand{\bs}[1]{p\!\!\!\!\! p \!\!\!\!\! p\!\!\!\!\! p}

\begin{document}
\title{Transport properties of the hot and dense sQGP}

\author{H.~Berrehrah$^1$, E.~Bratkovskaya$^1$, W.~Cassing$^2$ and R.~Marty$^{1,3}$}

\address{$^1$ Frankfurt Institute for Advanced Studies and Institute for Theoretical Physics, Johann Wolfgang Goethe Universit\"at, Ruth-Moufang-Strasse 1, 60438 Frankfurt am Main, Germany\\
$^2$ Institut f\"ur Theoretische Physik, Universit\"at Giessen, 35392 Giessen, Germany\\
$^3$ Institut de physique nucléaire de Lyon (IPNL), Universit\'e Claude Bernard Lyon 1, Villeurbanne, France}

\ead{berrehrah@fias.uni-frankfurt.de}

\begin{abstract}
The transport properties of the quark gluon plasma (QGP) are studied in a QCD medium at finite temperature and chemical potential. We calculate the shear viscosity $\eta(T,\mu_q)$ and the electric conductivity $\sigma_e(T,\mu_q)$ for a system of interacting massive and broad quasi-particles as described by the dynamical quasi-particle model ``DQPM'' at finite temperature $T$ and quark chemical potential $\mu_q$ within the relaxation time approximation. Our results are in a good agreement with lattice QCD at finite temperature and show clearly the increase of the transport coefficients with increasing $T$ and $\mu_q$. Our results provide the basic ingredients for the study of the hot and dense matter in the Beam Energy Scan (BES) at RHIC and CBM at FAIR.
\end{abstract}

\vspace*{-0.2cm}
\section{Introduction}

The exploration of the phase diagram of strongly interacting matter is a major field of modern high-energy physics. The transition from a hadronic to a partonic medium at small net-baryon densities is known to be a crossover. At high baryon densities one expects new phases of the strongly interacting matter. Besides the understanding of the thermodynamic properties of such systems, the transport properties at high temperature and density are of interest for many purposes. They are the key ingredients for hydrodynamic calculations ($\eta/s$, $\xi/s$) and transport simulations which compare the in and out-off equilibrium systems. They are also useful to confront different transport models, which study the expansion of the plasma created in relativistic heavy-ion collisions.

It is the purpose of this study to evaluate the transport coefficients for a system of interacting massive and broad quasi-particles as described by the dynamical quasi-particle model ``DQPM'' at finite temperature $T$ and quark chemical potential $\mu_q$ within the relaxation time approximation. The relaxation times for these particles are calculated using the transition rates corresponding to $qq$, $q\bar{q}$, $qg$, $gg$ elastic and $gg \leftrightarrow q \bar{q}$ inelastic scattering processes, where $q$ ($\bar{q}$) denotes the light (anti-)quark and $g$ the gluons of the sQGP.

\vspace*{-0.2cm}
\section{The DQPM at finite ($T$, $\mu_q$)}

 To evaluate the transition rates and cross sections of the partonic scattering processes in the sQGP using the leading order Born Feynman diagrams one needs to specify- in such a finite $T$ and $\mu_q$ medium- the values of the fundamental ingredients which are the coupling $\alpha_s$, the infrared regulator ``IR'' and the masses of external $q$,$\bar{q}$ and $g$ legs. These ingredients are provided by the DQPM in our calculations.
 
\begin{figure}[h!] 
\begin{minipage}{14.5pc}
\begin{center}
\includegraphics[width=14.8pc, height=14.3pc]{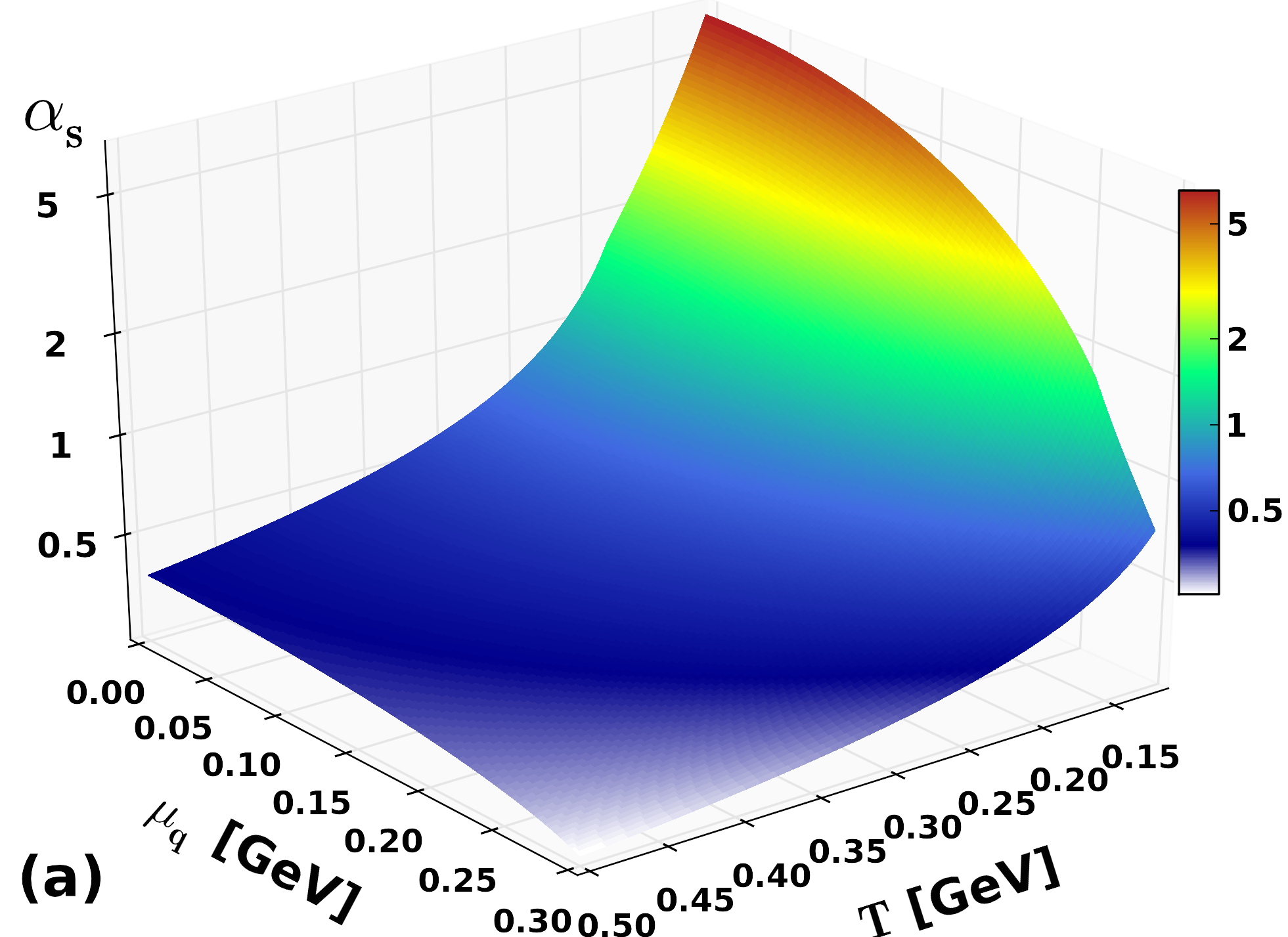}
\end{center}
\end{minipage}\hspace{0.3pc}
\begin{minipage}{14.5pc}
\begin{center}
\includegraphics[width=14.8pc, height=14.3pc]{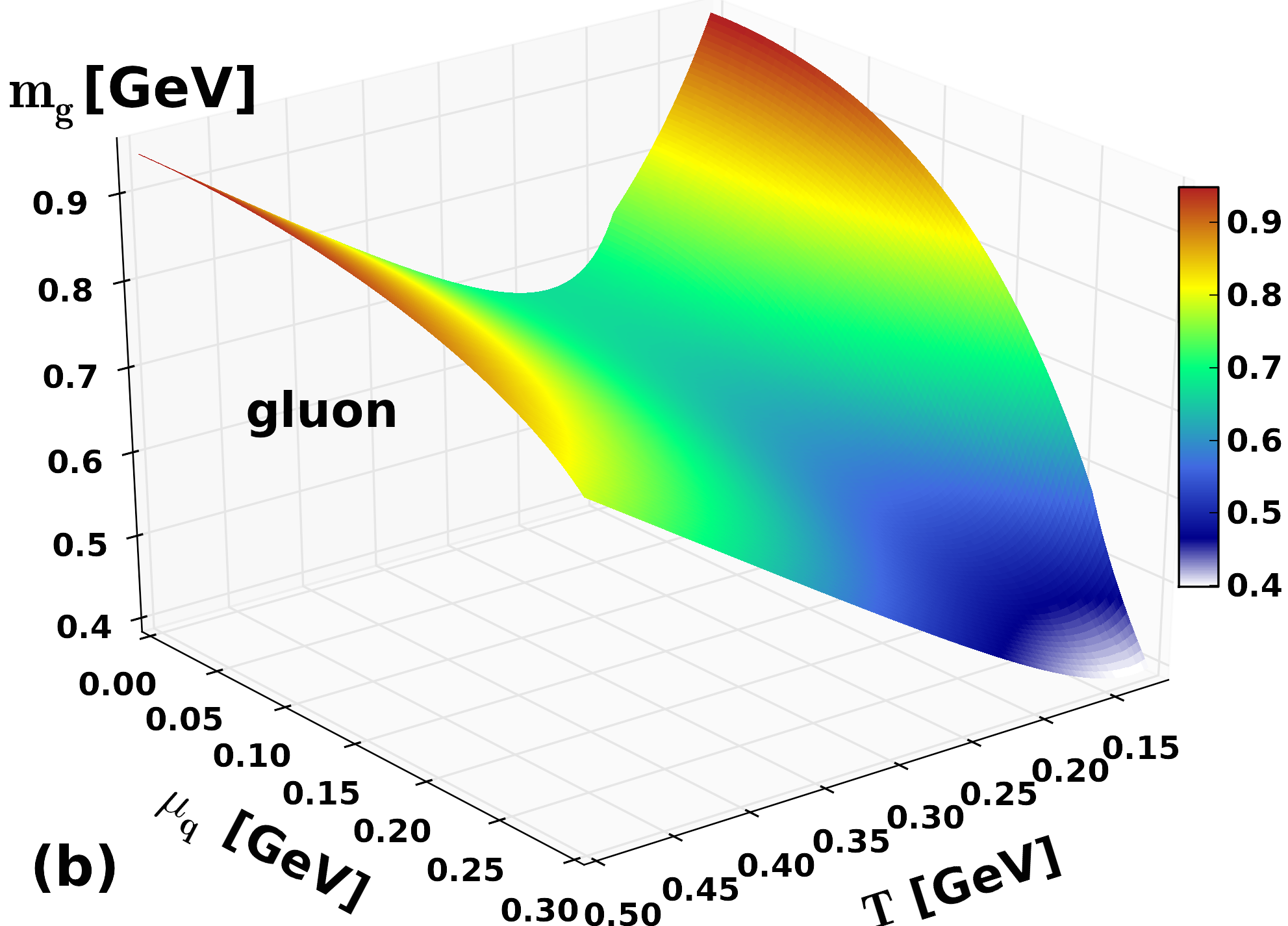}
\end{center}
\end{minipage}\hspace{1pc}
\begin{minipage}{7.2pc}
\begin{center}
\caption{\emph{(Color online) The DQPM coupling constant $\alpha_s$ (a) and gluon mass (b) as a function of the temperature $T$ and quark chemical potential $\mu_q$.}}
\label{fig:alphaMassgDQPM}
\end{center}
\end{minipage}
\end{figure}

 The DQPM coupling constant and gluon mass (which is considered also as IR) are shown in figure \ref{fig:alphaMassgDQPM} (a) and (b) as a function of $T$ and $\mu_q$. One sees that $\alpha_s$ is much larger than one near $T_c (\mu_q)$; the non-perturbative effects then become most apparent at these temperatures. The increase of $\mu_q$ leads to a decrease of $\alpha_s$ and $m_{q,\bar{q},g}$. The DQPM provides also the particle width $\gamma$ and the Breit-Wigner spectral function $\rho^{BW}$ since the constituents of the sQGP are considered as strongly interacting massive effective quasiparticles \cite{Wcassing2009EPJS}. 
\vspace*{-0.2cm}
 
\vspace*{-0.6cm}
\section{$g$, $q$ and $\bar{q}$ scattering at finite ($T$, $\mu_q$)}

  By dressing the quark and gluon lines with the effective propagators at finite temperature and chemical potential, we derive the on- and off-shell cross sections and transition rates for the dominant partonic reactions in the sQGP. The partons are dressed by effective DQPM pole masses in the on-shell case and are dressed by the DQPM spectral functions with a finite width in the off-shell case. We refer to the on-shell study by DpQCD approach (Dressed pQCD) and to the off-shell case by IEHTL approach (Infrared Enhanced Hard Thermal Loop) \cite{BerrehrahPubli,Berrehrah:2014tva}. 
    
\begin{figure}[h!] 
\begin{minipage}{14.5pc}
\begin{center}
\includegraphics[width=14.7pc, height=14.2pc]{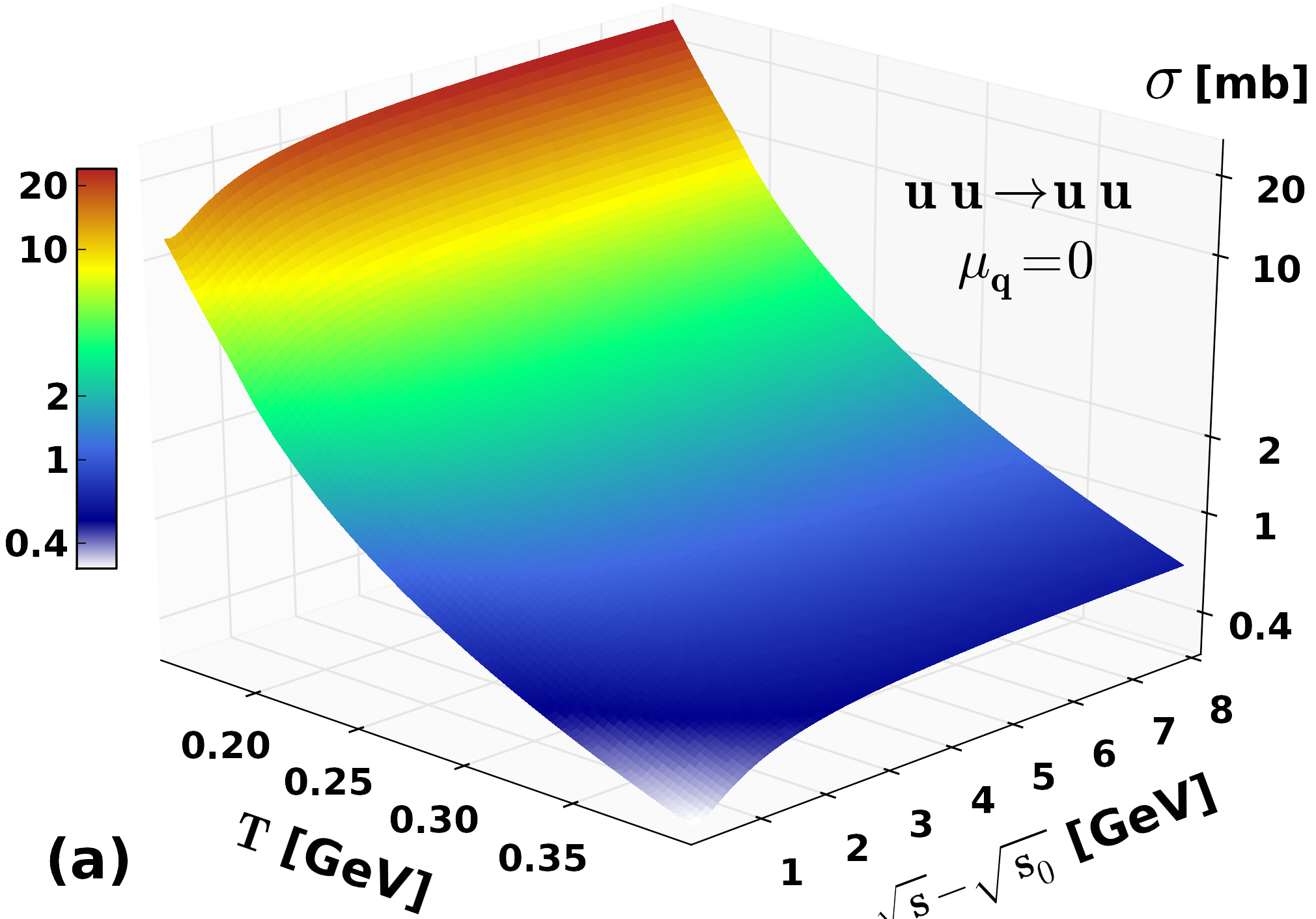}
\end{center}
\end{minipage}\hspace{0.3pc}
\begin{minipage}{14.5pc}
\begin{center}
\includegraphics[width=14.7pc, height=14.2pc]{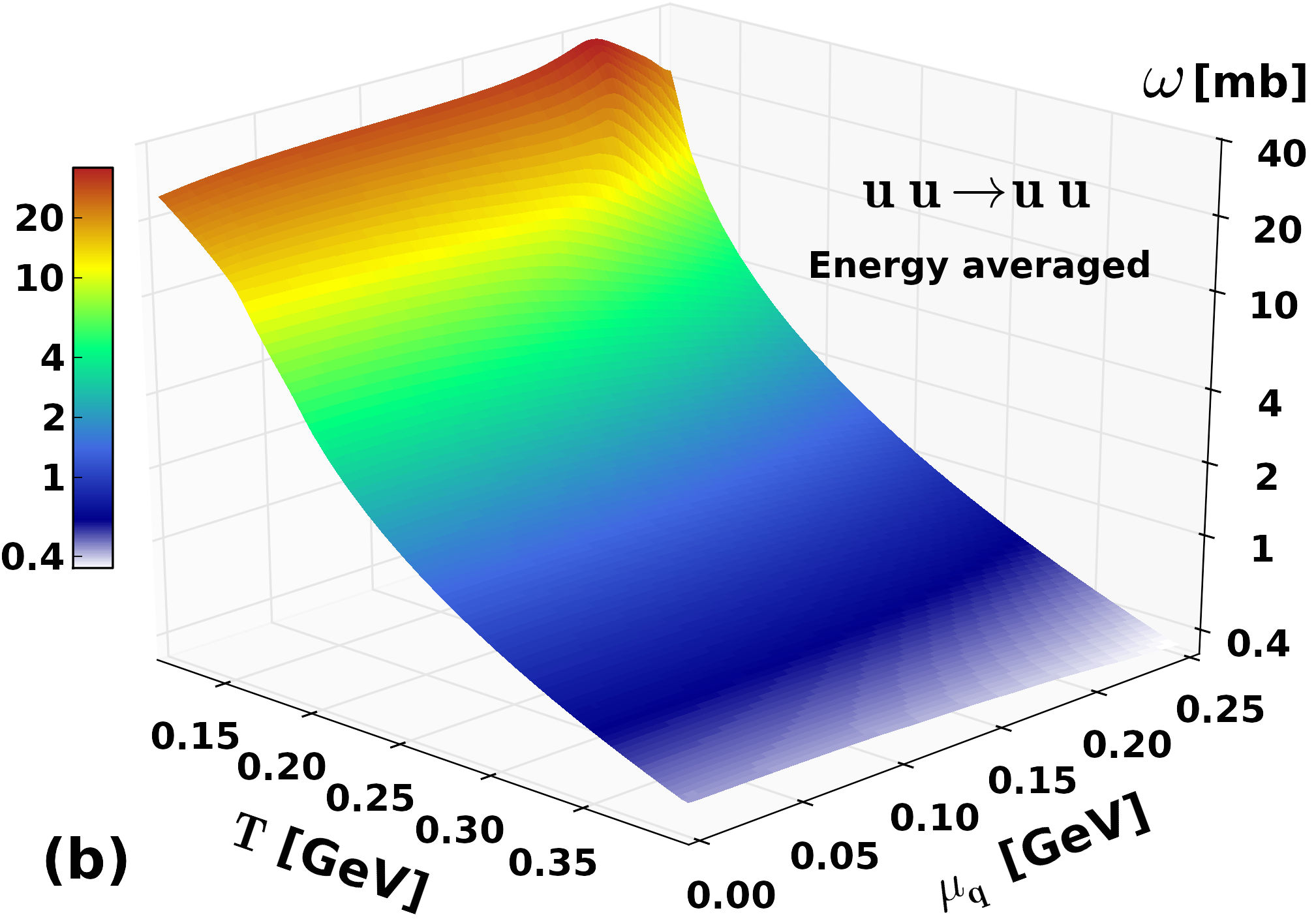}
\end{center}
\end{minipage}\hspace{1pc}
\begin{minipage}{7.2pc}
\begin{center}
\caption{\emph{(Color online) $u \ u \rightarrow u \ u$ cross section as a function of $\sqrt{s}-\sqrt{s}_0$, with $s_0$ denoting the threshold (a) and thermal transition rate as of function of $(T, \mu_q)$ (b).}}
\label{fig:SigmaOmegauuDpQCD}
\end{center}
\end{minipage}
\end{figure}

  The effect of the spectral function in the off-shell approach is demonstrated to be small because of the small DQPM width \cite{BerrehrahPubli,Berrehrah:2014tva}. However the off-shell IEHTL calculation shows a reduction of the kinematic threshold as compared to the DpQCD on-shell non-perturbative approach. Such an effect leads only to a small shift in the partonic transition rates. Therefore, we show in Figs.\ref{fig:SigmaOmegauuDpQCD} (a) and (b) the DpQCD $u \ u \rightarrow u \ u$ cross section $\sigma$ as a function of $T$ and $\sqrt{s} - \sqrt{s_0}$ ($s_0$ is the energy threshold) and transition rate $\omega$ as a function of $T$ and $\mu_q$, respectively.
\vspace*{-0.2cm}
 
  Figs.\ref{fig:SigmaOmegauuDpQCD} (a) and (b) show a large enhancement of $\sigma$ and $\omega$ for $T$ close to $T_c(\mu_q)$ due to the infrared enhancement of $\alpha_s(T, \mu_q)$ as seen in Fig.\ref{fig:alphaMassgDQPM}. Despite the lower value of the IR regulator at higher values of $\mu_q$ the smaller value of $\alpha_s$ at finite $\mu_q$ explains the decrease of $\sigma$ and $\omega$ at finite and large $\mu_q$.
  
  The thermal transition rate $\omega$ can be parametrized by a power law in $T$ for each value of $\mu_q$. Having almost the same power laws at finite $\mu_q$ as compared to $\mu_q = 0$ for temperatures larger than $T_c(\mu_q = 0)$, we observe some scaling effects in the transport coefficients at these temperatures.
  

\vspace*{-0.2cm}
\section{sQGP transport properties at finite ($T$, $\mu_q$)}

 We compute transport coefficients using the relaxation time approximation. 
 In the dilute gas approximation the relaxation time $\tau$ is obtained for on-shell particles $(\tau_c^{-1})_{\textrm{DpQCD}}$ and for off-shell quasi-particles $(\tau_c^{-1})_{\textrm{IEHTL}}$,$(\tau_c^{-1})_{\textrm{DQPM}}$ by Eq.(\ref{equ:3}) \cite{Marty:2013ita}, where $n_{i}$ is the $q$/$\bar{q}$ or $g$ density.  For the DQPM we do not need the explicit cross sections since the inherent quasi-particle width $\gamma_i (T)$ directly provides the total interaction rate \cite{Wcassing2009EPJS}.

\vspace{-0.7cm}
{\setlength\arraycolsep{-1pt}
\begin{eqnarray}
\label{equ:3}
& & \displaystyle(\tau_i^{-1})_{\textrm{DpQCD}}\! = \!\!\!\!\!\!\! \sum\limits_{j \in{q,\bar{q},g}} \!\!\!\!\!\! n_j^{\textrm{on}} (T, \mu_j) \omega_{\textrm{j i}}^{DpQCD} (T, \mu_j), \hspace{0.15cm} \displaystyle n_j^{\textrm{on}} (T, m_j,\mu_j) \!\!=\!\!\!\! \int \!\!\! \frac{d^3 p}{(2 \pi)^3} f_j (p, T, m_j,\mu_j), \hspace{0.15cm} \displaystyle (\tau_i^{-1})_{DQPM} \!=\!\! \frac{\hbar c}{\gamma_i (T,\mu_i)}
\nonumber\\
& & {} \displaystyle(\tau_i^{-1})_{\textrm{IEHTL}} \!\!= \!\!\!\!\! \sum\limits_{j \in{q,\bar{q},g}}\!\!\! n_j^{\textrm{off}} (T,\mu_j) \ \omega_{\textrm{j i}}^{IEHTL} (T,\mu_j), \hspace{0.4cm} \displaystyle n_j^{\textrm{off}} (T,\mu_j) \!\!=\!\!\! \int \!\!\!\! \int \!\! \frac{d^3 p}{(2 \pi)^3} \rho_j^{BW} (m_j) d m_j \ f_j (p, T, m_j,\mu_j).
\end{eqnarray}}
\vspace{-0.24cm}

\begin{figure}[h!] 
\begin{minipage}{14.5pc}
\begin{center}
\includegraphics[width=14.7pc, height=14.4pc]{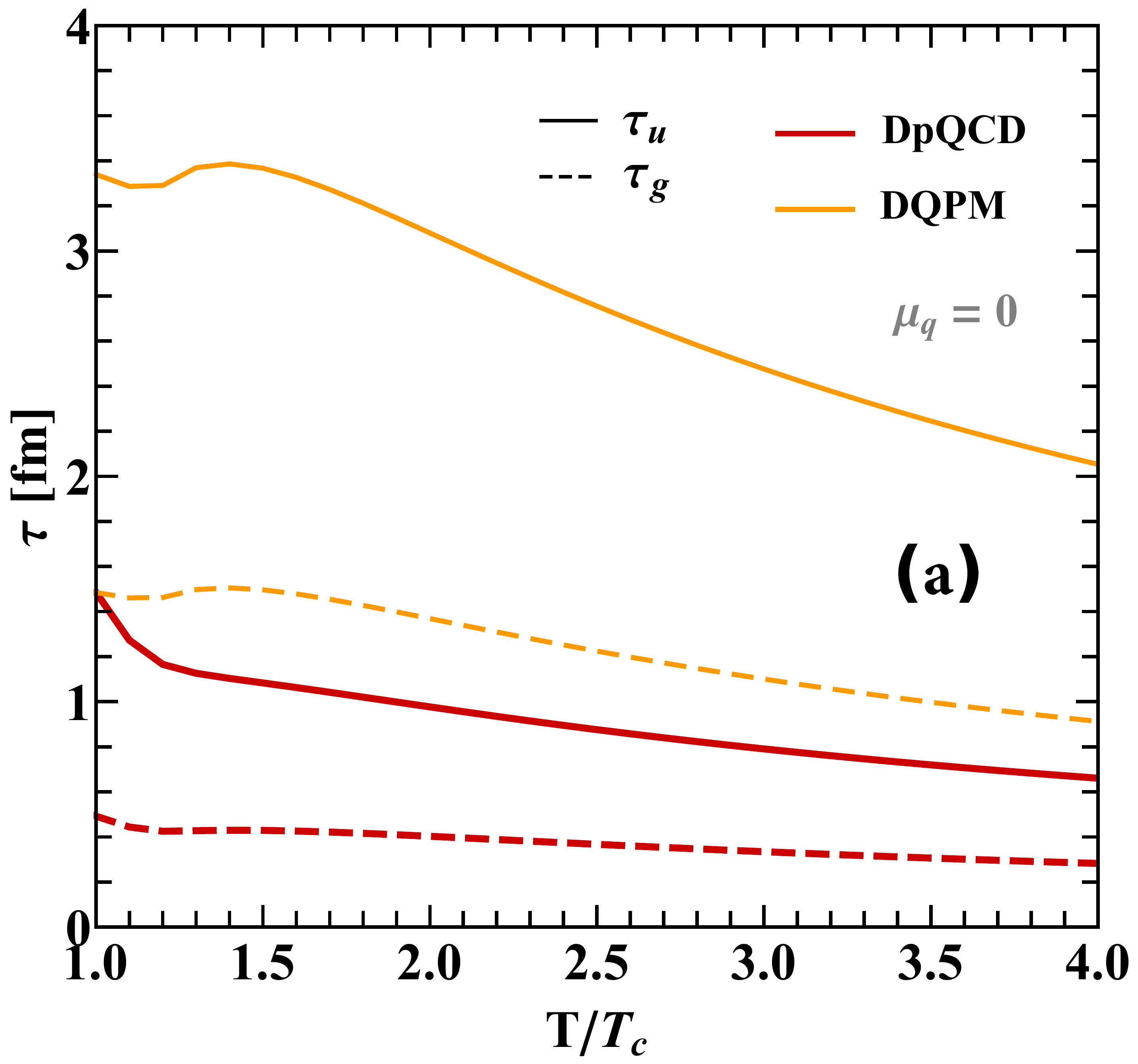}
\end{center}
\end{minipage}\hspace{0.3pc}
\begin{minipage}{14.5pc}
\begin{center}
\includegraphics[width=14.7pc, height=14.4pc]{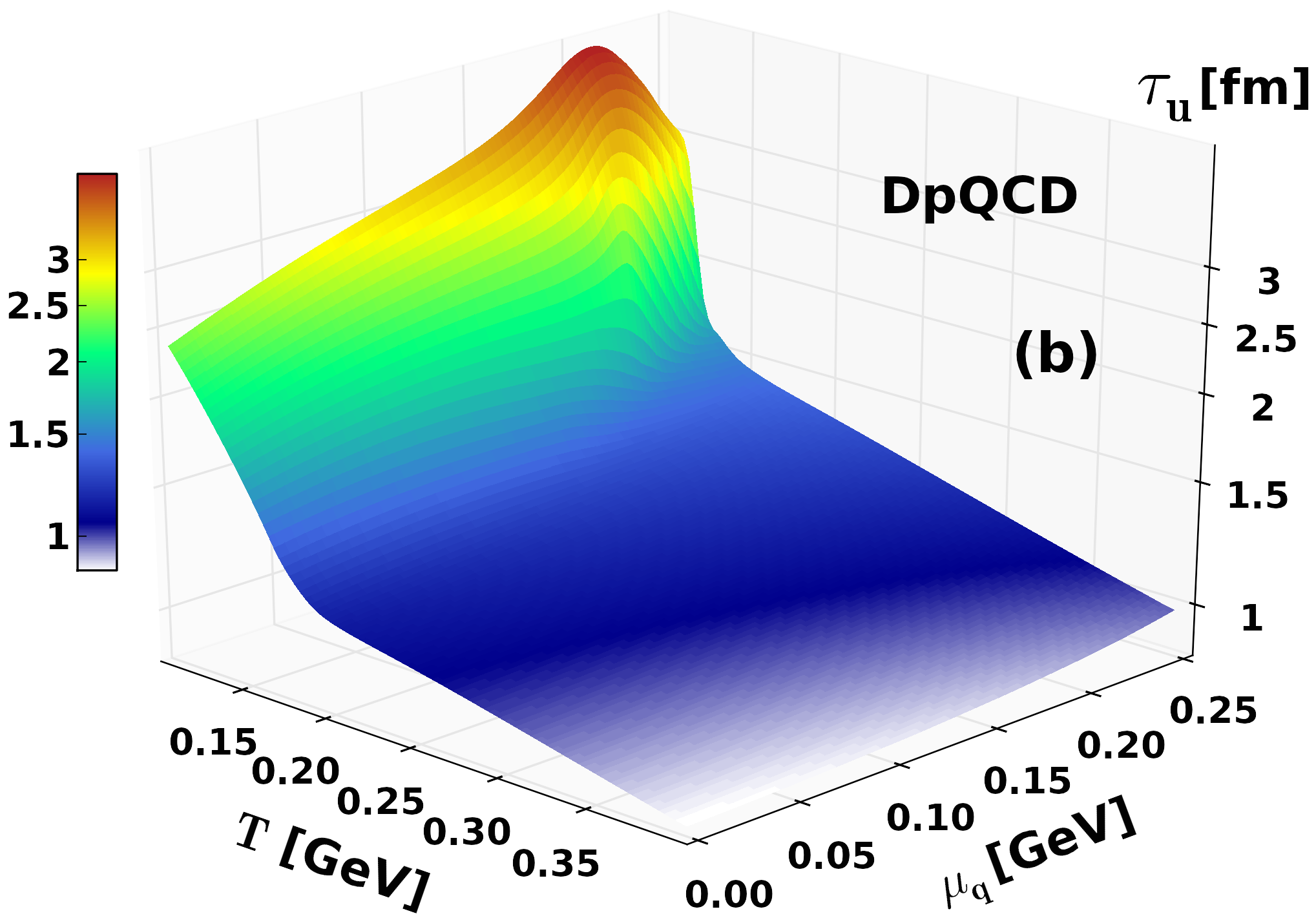}
\end{center}
\end{minipage}\hspace{1pc}
\begin{minipage}{7.2pc}
\begin{center}
\caption{\emph{(Color online) $u$-quark and gluon relaxation times $\tau_{u,g}$ following the on- and off-shell models as a function of temperature $T$ for $\mu_q =0$ (a) and $\tau_{u}$ given by DpQCD as a function of $(T,\mu_q)$ (b).}}
\label{fig:tau}
\end{center}
\end{minipage}
\end{figure}

  Figs. \ref{fig:tau}-(a) and (b) show the $u$-quark and gluon relaxation times for on- and off-shell partons at finite $T$ and finite $(T,\mu_q)$, respectively. Figs. \ref{fig:tau}-(a) shows that $\tau$ decreases with temperature since the quark and gluon densities are increasing functions of temperature. We can evaluate -in terms of powers of $T$- the behavior of $\tau_{u,g}$ for the different approaches. The higher power coefficients in the relaxation time for $T < 1.2 T_c$ in DpQCD/IEHTL approaches can be traced back to the infrared enhancement of the effective coupling. A finite $\mu_q$ increases smoothly $\tau_u$ especially close to $T_c (\mu_q)$, as shown in Fig. \ref{fig:tau}-(b)
\vspace*{-0.2cm}

\begin{figure}[h!] 
\begin{minipage}{14.5pc}
\begin{center}
\includegraphics[width=14.9pc, height=14.5pc]{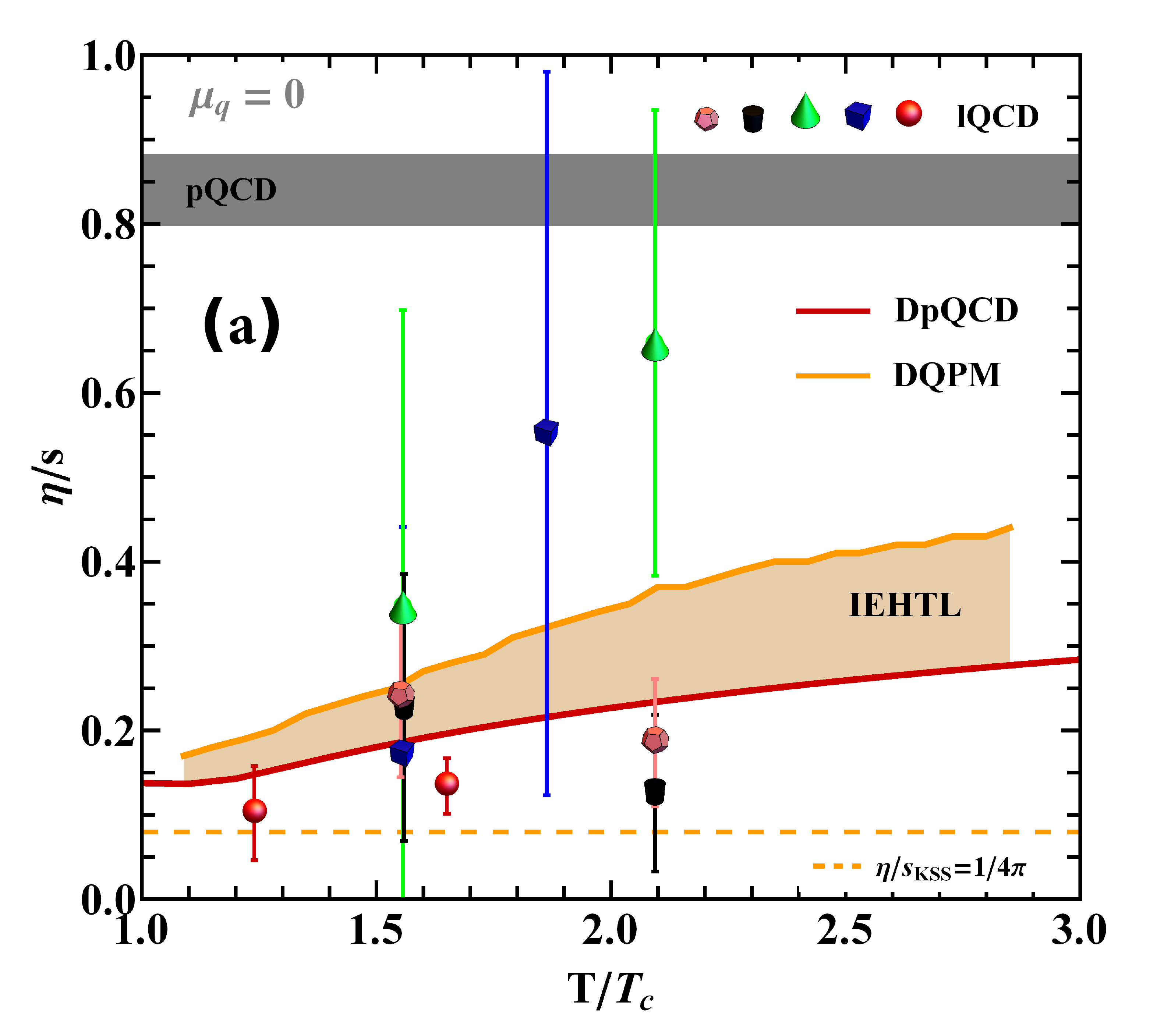}
\end{center}
\end{minipage}\hspace{0.3pc}
\begin{minipage}{14.5pc}
\begin{center}
\includegraphics[width=14.9pc, height=14.5pc]{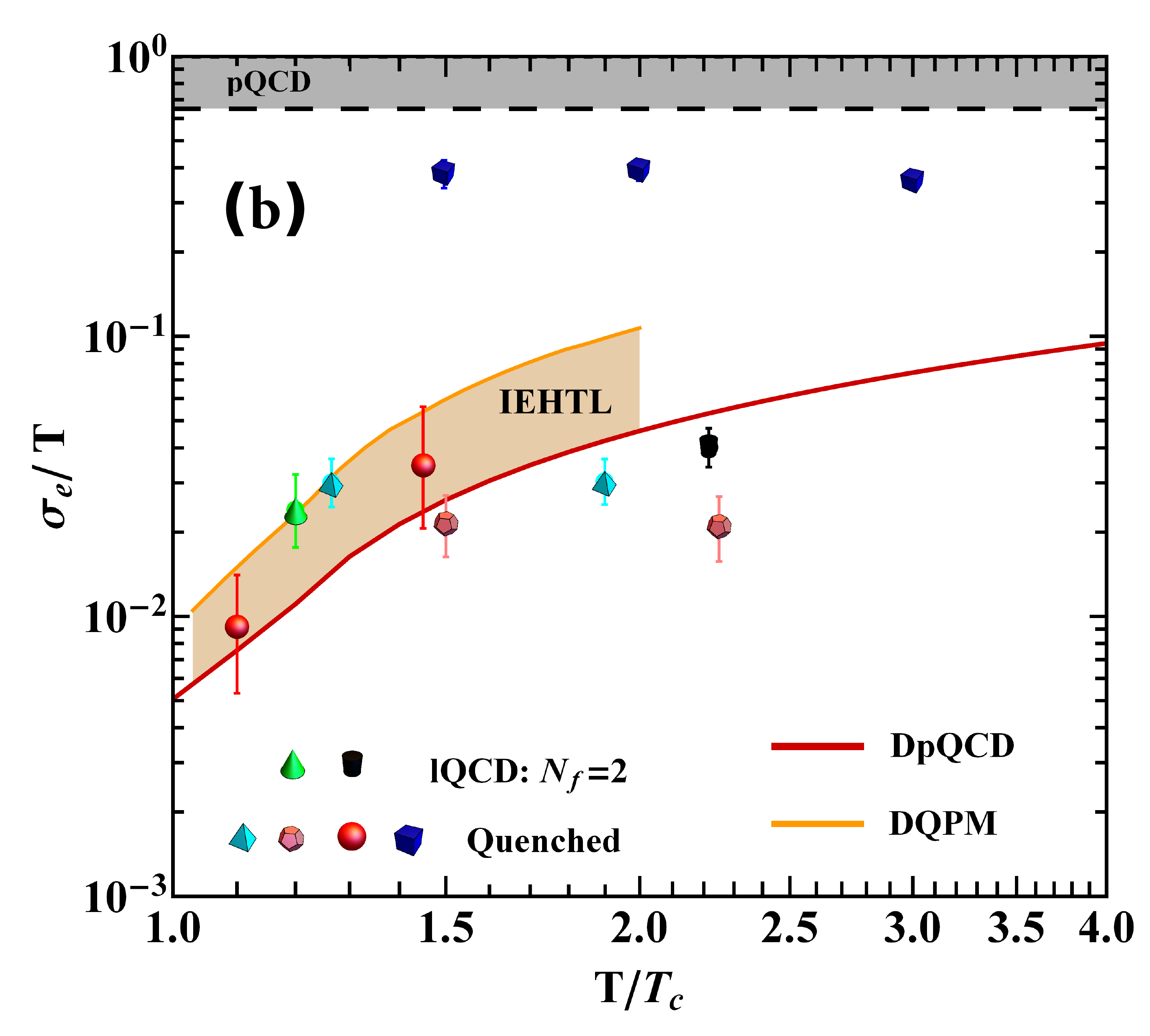}
\end{center}
\end{minipage}\hspace{1pc}
\begin{minipage}{7.2pc}
\begin{center}
\caption{\emph{(Color online) $\eta/s$ (a) and $\sigma_e/T$ (b) following the on- and off-shell models as a function of temperature $T$ for $\mu_q =0$. The lattice QCD data are given from \cite{lQCDdata}.}}
\label{fig:EtaSigmae}
\end{center}
\end{minipage}
\end{figure}

  Using the relaxation times $\tau$ shown above we compute the different transport coefficients of the sQGP. As an example we show the shear viscosity $\eta$ and electric conductivity $\sigma_e$ as a function of $(T,\mu_q)$ following our models where $\eta$ and $\sigma_e$ are given (for the case of on-shell partons) by
\vspace{-0.3cm}
{\setlength\arraycolsep{-1pt}
\begin{eqnarray}
\label{equ:2}
& & \eta^{\textrm{on}}(T,\mu_q) = \frac{1}{15 T} g_g \!\!\int\!\! \frac{d^3 p}{(2\pi)^3} \tau_g^{\textrm{on}} f_g \frac{{\bf p}^4}{E_g^2} + \frac{1}{15 T} \frac{g_q}{6} \!\!\int\!\! \frac{d^3 p}{(2\pi)^3} \Bigg[ \sum_q^{u,d,s} \tau_q^{\textrm{on}} f_q + \sum_{\bar q}^{\bar u,\bar d,\bar s} \tau_{\bar q}^{\textrm{on}} f_{\bar q} \Bigg] \frac{{\bf p}^4}{E_q^2},
\nonumber\\
& & {} \sigma_e^{\textrm{on}} (T,\mu_q) = \sum_{f,\bar{f}}^{u,d,s} \frac{e_f^2}{m_f (T, \mu_q)} n_f^{\textrm{on}} (T, \mu) \tau_f^{\textrm{on}} (T, \mu_q),
\end{eqnarray}}
\vspace{-0.3cm}
  
with $e_q$ the electric charge of quarks. For off-shell partons Eqs.(\ref{equ:2}) can be generalized by taking into account the parton spectral functions, off-shell relaxation times and parton densities. Figs.\ref{fig:EtaSigmae}-(a) and (b) show that $\eta/s$, where $s$ is the DQPM entropy density, and $\sigma_e/T$ given by DpQCD/IEHTL and DQPM models are in the range of the lQCD data. On the other side --going from pQCD to non-pertrubative based models-- leads to a reduction of $\eta/s$ and $\sigma_e/T$. As a function of temperature $\eta/s$ shows a minimum around $T_c$ and then increases slowly for higher temperatures. $\sigma_e/T$ increases slowly until it reaches at high temperatures the pQCD regime where $\sigma_e/T \sim const$.   
\vspace*{0.05cm}

\begin{figure}[h!] 
\begin{minipage}{14.5pc}
\begin{center}
\includegraphics[width=14.7pc, height=14.3pc]{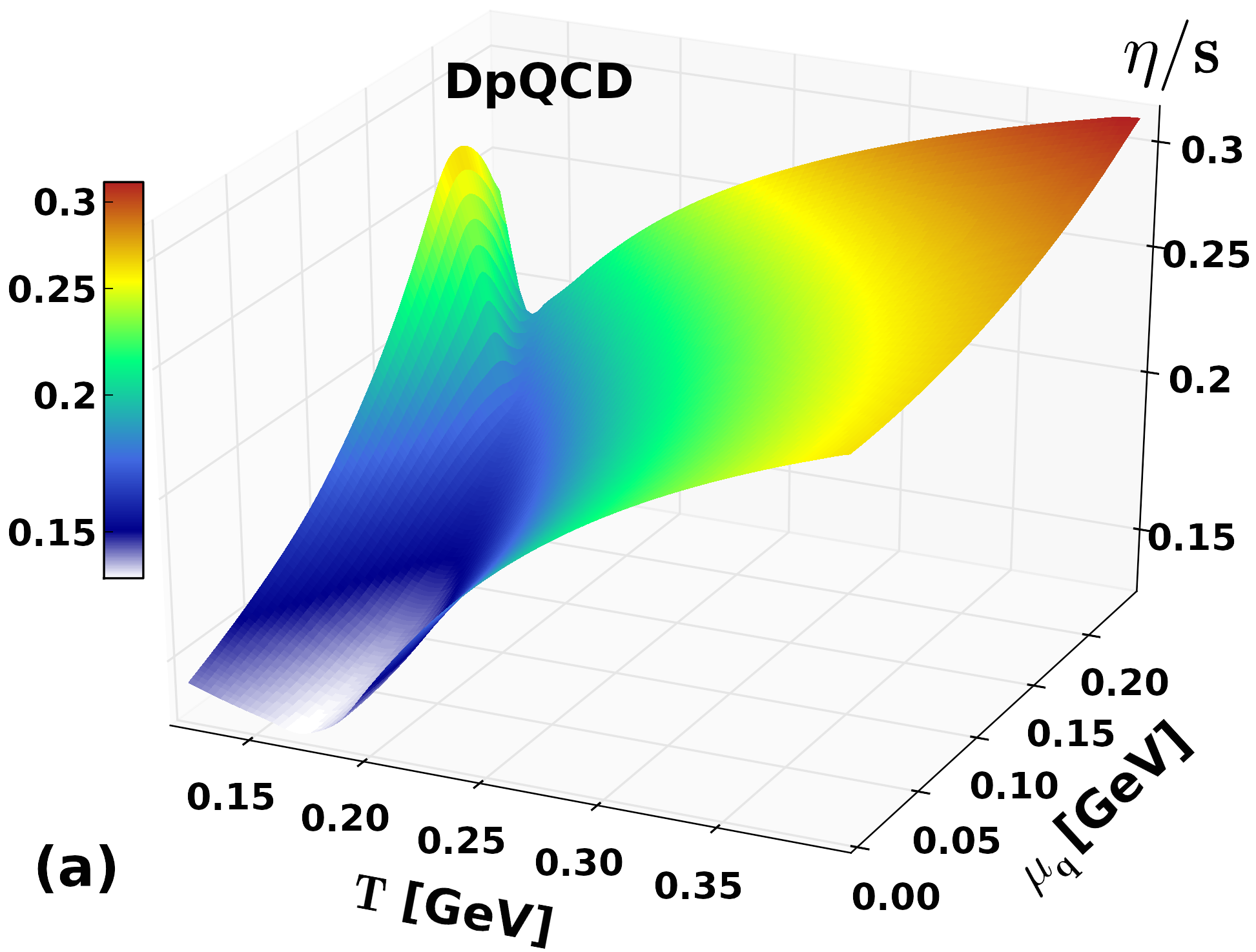}
\end{center}
\end{minipage}\hspace{0.3pc}
\begin{minipage}{14.5pc}
\begin{center}
\includegraphics[width=14.7pc, height=14.3pc]{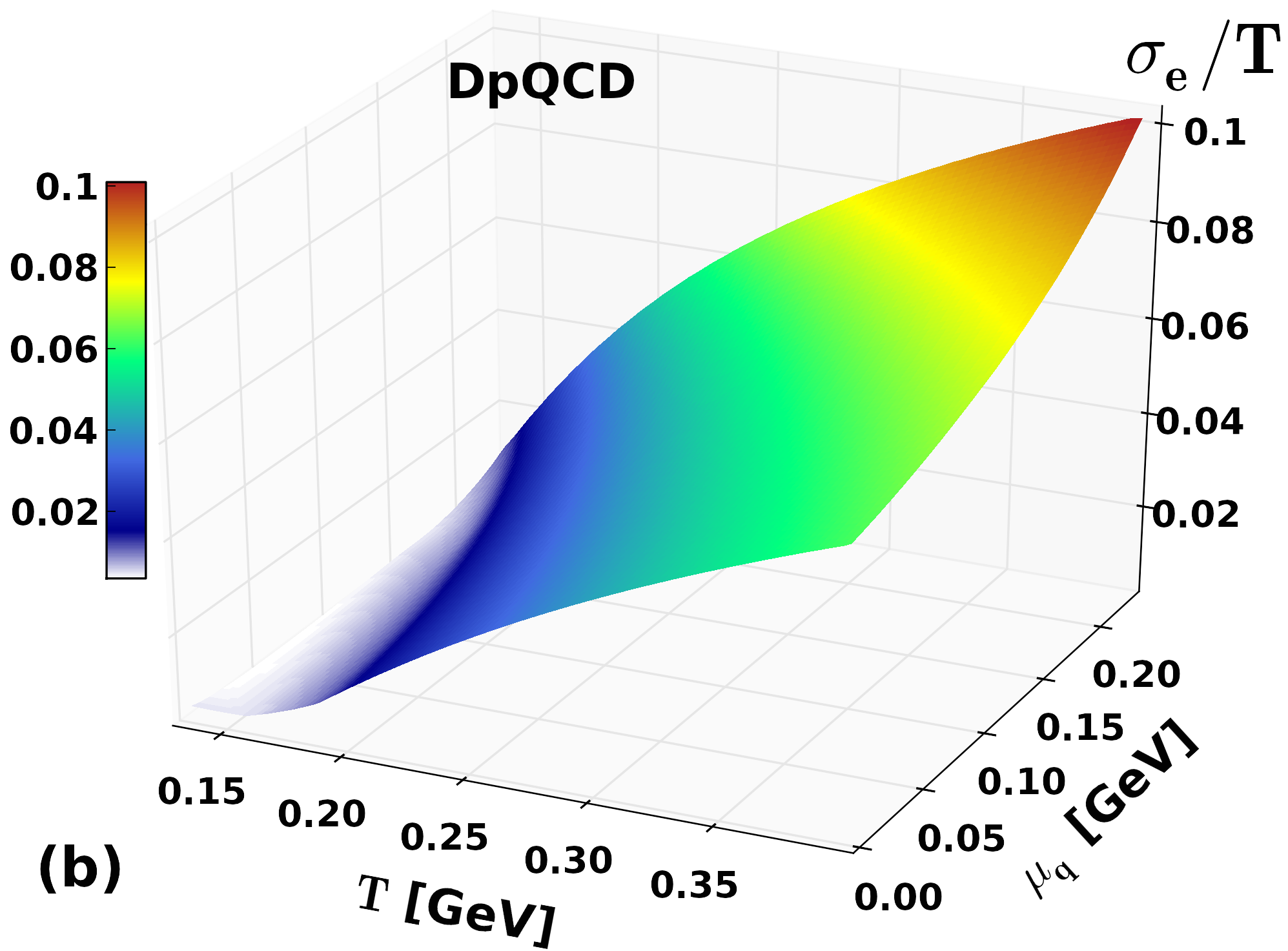}
\end{center}
\end{minipage}\hspace{1pc}
\begin{minipage}{7.2pc}
\begin{center}
\caption{\emph{(Color online) $\eta/s$ (a) and $\sigma_e/T$ (b) following the on-shell DpQCD model as a function of $(T, \mu_q)$.}}
\label{fig:EtaSigmaeTmu}
\end{center}
\end{minipage}
\end{figure}

Finally the $(T,\mu_q)$ dependencies of $\eta/s$ and $\sigma_e/T$ are shown in Figs.\ref{fig:EtaSigmaeTmu}-(a) and (b) respectively. One sees that $\eta$ increases like $T^{3}$ for large temperature for all chemical potentials in such a way that $\eta/s$ is almost constant. $\eta/s$ and $\sigma_e/T$ show a smooth increase as a function of $(T, \mu_q)$.
\vspace*{-0.18cm}

\vspace*{-0.7cm}
\section{Summary} 

  We have presented a microscopic computation of sQGP transport properties at finite temperature and chemical potential using the relaxation times of the constituents of the system. The partons are considered as strongly interacting massive quasi-particles described by the DQPM \cite{Wcassing2009EPJS,BerrehrahPubli,Berrehrah:2014tva} in which quarks and gluons have a finite mass and width that vary with $T$ and $\mu_q$. We have demonstrated that our $\eta/s$ and $\sigma_e/T$ are in the range of lQCD data. The transport coefficients in our models show a smooth dependence on $(T,\mu_q)$. This is consistent with a crossover transition at small $\mu_q$ in line with lattice calculations.

\vspace*{-0.2cm}
\section*{References}
\bibliographystyle{iopart-num}
\bibliography{References}
 
\end{document}